\documentclass[%
 reprint,
superscriptaddress,
aps,
prd
]{revtex4-2}
\usepackage{amsmath} 
\usepackage{graphicx}% Include figure files
\usepackage{dcolumn}% Align table columns on decimal point
\usepackage{bm}% bold math
\usepackage{hyperref}% add hypertext capabilities
\begin{document}

\preprint{APS/123-QED}

% \title{Manuscript Title:\\with Forced Linebreak}% Force line breaks with \\
%\title{Self-Tuned $\Lambda$ Inflation: elegant Exit from a Bare Cosmological Constant}
%\title{Self-tuned $\Lambda$ inflation: an elegant solution to the cosmological constant problem}
\title{Inflation driven by a bare cosmological constant and its graceful exit}

% \thanks{A footnote to the article title}%

\author{Chengsheng Mu}
\affiliation{School of Physics and Astronomy, Beijing Normal University, Beijing 100875, China;}
\affiliation{Institute for Frontiers in Astronomy and Astrophysics, Beijing Normal University, Beijing 102206, China}%

\author{Shuxun Tian}
\email{tshuxun@bnu.edu.cn}
\affiliation{School of Physics and Astronomy, Beijing Normal University, Beijing 100875, China;}

\author{Shuo Cao}
\email{caoshuo@bnu.edu.cn}
\affiliation{School of Physics and Astronomy, Beijing Normal University, Beijing 100875, China;}
\affiliation{Institute for Frontiers in Astronomy and Astrophysics, Beijing Normal University, Beijing 102206, China}%

\author{Zong-Hong Zhu}
\email{zhuzh@bnu.edu.cn}
\affiliation{School of Physics and Astronomy, Beijing Normal University, Beijing 100875, China;}
\affiliation{Institute for Frontiers in Astronomy and Astrophysics, Beijing Normal University, Beijing 102206, China}%

\date{\today}% It is always \today, today,
             %  but any date may be explicitly specified

\begin{abstract}

Vacuum energy, a prediction of quantum field theory, manifests itself as a cosmological constant in general relativity. In this Letter, we propose a novel inflationary scenario driven by a bare cosmological constant $\Lambda$, which terminates naturally through a self-tuning mechanism. Within Fab-Four gravity, self-tuning destabilizes the de Sitter state and drives the system toward a stiff-fluid attractor, thereby yielding a graceful exit. We construct two explicit models in which the slow-roll parameter evolves exponentially or as a power law. We show that the latter model, derived from center-manifold dynamics, significantly relaxes the required tuning of initial conditions. Our results establish, for the first time, that bare-vacuum-energy inflation with natural termination constitutes a viable dynamical possibility.

\end{abstract}

%\keywords{Suggested keywords}%Use showkeys class option if keyword
                              %display desired
\maketitle

\section{Introduction}
Vacuum energy is typically discussed as a problem in modern cosmology, yet it may also serve as a resource. In quantum field theory it is unavoidable, yet its observed gravitational imprint is tiny compared with the natural particle-physics expectation, yielding the cosmological constant problem \cite{carrollCosmologicalConstant2001,1989RvMP...61....1W,martinEverythingYouAlways2012}. If a large bare vacuum energy existed in the early universe, the conservative question is whether it could have driven inflation before being dynamically screened. The appeal is similar to Higgs inflation \cite{bezrukovStandardModelHiggs2008}, where one uses the experimentally established Higgs field -- albeit with a nonminimal coupling to gravity -- rather than introducing a new inflation sector \cite{guthInflationaryUniversePossible1981,1982PhLB..108..389L,PhysRevLett.48.1220}. The analogous idea explored in this Letter is to use another ingredient already present in the theory: vacuum energy itself. 
The main obstacle is achieving a graceful exit. A strictly constant $\Lambda$ supports a de Sitter phase, but it does not decay, so inflation persists indefinitely. By contrast, classic inflationary scenarios were built with exit mechanisms, such as scalar slow roll \cite{1982PhLB..108..389L,PhysRevLett.48.1220,liddleFormalizingSlowrollApproximation1994}. A viable $\Lambda$-driven picture must therefore do two things at once. It must realize a quasi-de Sitter stage from a large bare cosmological constant and then suppress its gravitational effect quickly enough to end inflation. Self-tuning scalar-tensor theories provide a natural setting for this question. Weinberg's no-go argument rules out adjustment mechanisms when the vacuum is kept fully Poincar\'{e} invariant and all fields settle to constant values; in that case, canceling an arbitrary vacuum energy still requires fine tuning \cite{1989RvMP...61....1W}. 

Self-tuning evades this logic by allowing the scalar to remain nontrivial on the vacuum solution, typically through a time-dependent background \cite{charmousisGeneralSecondOrderScalarTensor2012,charmousisSelftuningDerivationFab2012,copelandCosmologyFabFour2012}. The Fab-Four theory is a well-known realization of this mechanism within Horndeski gravity \cite{charmousisGeneralSecondOrderScalarTensor2012,charmousisSelftuningDerivationFab2012,copelandCosmologyFabFour2012}. The previous studies of self-tuning, however, have emphasized late-time screening or asymptotic self-tuned vacua, rather than transient early-universe dynamics in which a large bare cosmological constant first drives inflation and is only subsequently screened. In this Letter, we propose a novel inflationary scenario driven by a bare cosmological constant $\Lambda$, which terminates naturally via a self-tuning mechanism. Within Fab-Four gravity, we construct two explicit models in which the de Sitter state sourced by a bare $\Lambda$ is an unstable critical point rather than the final attractor. In Model I, the slow-roll parameter grows exponentially. In Model II, the de Sitter point is non-hyperbolic, so trajectories are rapidly attracted to a center manifold and then evolve slowly along it, leading to a power-law growth of the slow-roll parameter and weaker sensitivity to initial conditions. In both cases the system evolves toward a stiff-fluid attractor, showing that inflation driven by a bare cosmological constant can end through a graceful exit rather than eternal de Sitter expansion.
Our results demonstrate, for the first time, the existence of this mechanism at the proof-of-principle level.

\section{Model and physical implications} The Fab-Four theory is composed of four Lagrangians $\mathcal{L}_i = \sqrt{-g} L_i$, where the scalar sectors are $L_1 = V_1 G^{\mu\nu} \nabla_\mu\phi \nabla_\nu\phi$, $L_2 = V_2 P^{\mu\nu\alpha\beta} \nabla_\mu\phi \nabla_\alpha\phi \nabla_\nu \nabla_\beta\phi$, $L_3 = V_3 R$, and $L_4 = V_4 \hat{G}$ \cite{charmousisGeneralSecondOrderScalarTensor2012}. In this notation, $G^{\mu\nu}$ denotes the Einstein tensor, $P^{\mu\nu\alpha\beta}$ is the double dual of the Riemann tensor, and $\hat{G}$ represents the Gauss-Bonnet topological invariant, while the coefficients $V_i(\phi)$ denote arbitrary potential functions of the scalar field.
\subsection{Exponential slow-roll parameter}
Adhering to the Brans-Dicke philosophy of avoiding extrinsic mass scales \cite{bransMachsPrincipleRelativistic1961}, we assign the scalar field the dimension of length. Taking $\mathcal{L}_3$ as the standard Einstein-Hilbert term and assuming polynomial couplings, dimensional analysis uniquely dictates the potentials as $V_{1}(\phi)=\alpha_1$, $V_2(\phi)=\alpha_2\phi$, and $V_4(\phi)=\alpha_4 \phi^2$.

In a vacuum-dominated spatially flat FLRW background, varying the action yields the modified Friedmann and scalar field equations \footnote{While intrinsic spatial curvature ($k \neq 0$) formally alters the asymptotic structure of the critical point, the inflationary attractor nature ensures that $\Omega_k$ decays exponentially. Thus, the derived trajectories represent the physical asymptotic behavior of the system.}. To characterize the global evolution where phase space variables may diverge, we employ Poincaré compactification \cite{meissDifferentialDynamicalSystems2007}. This projects the unbounded phase space onto a unit hemisphere via the coordinate transformation $y_1 = x_1 / \sqrt{1 + x_1^2 + x_2^2}$ and $y_2 = x_2 / \sqrt{1 + x_1^2 + x_2^2}$, allowing stability analysis of the equatorial boundary to rigorously identify late-time global attractors. The efficacy of Poincaré compactification hinges on selecting variables that capture the system's asymptotic scaling. Motivated by power-law solutions implying $\dot{\phi} \propto H \phi$, we define $x_1 = {\dot{\phi}}/{c}$, $x_2 = {\phi H}/{c}$, and $\Omega_\Lambda = {\Lambda c^2}/{H^2}$. Crucially, this normalization ensures that $x_1$ and $x_2$ remain of the same order as they diverge, thereby facilitating the resolution of the dynamics at infinity. Adopting $N \equiv \ln a$ as time, the background equations are recast into an autonomous system satisfying
\begin{subequations}
\label{eq:coupled_system}
\begin{align}
    \frac{\mathrm{d} x_1}{\mathrm{d} N} &= \frac{x_1 (x_1^2 \alpha_2 - \alpha_1)}{\alpha_1 - 3 x_1 x_2 \alpha_2} \nonumber \\
    &\quad + \frac{(1+\gamma) [x_2 (9 x_1^2 \alpha_2 + 16 \alpha_4) - 4 x_1 \alpha_1]}{2 (\alpha_1 - 3 x_1 x_2 \alpha_2)}, \label{eq:dx1} \\[1ex]
    \frac{\mathrm{d} x_2}{\mathrm{d} N} &= x_1 + \gamma x_2. \label{eq:dx2}
\end{align}
\end{subequations}
The system is closed by the auxiliary variable $\gamma$, which depends on the phase space coordinates through the rational function $\gamma \equiv \mathrm{d} lnH/\mathrm{d} N$.

\begin{figure}[t] % [b] 表示尽量放在底部，也可以用 [t] 放在顶部
    % width=\columnwidth 是关键，它让图片宽度自动等于当前分栏的宽度
    \includegraphics[width=0.85\columnwidth]{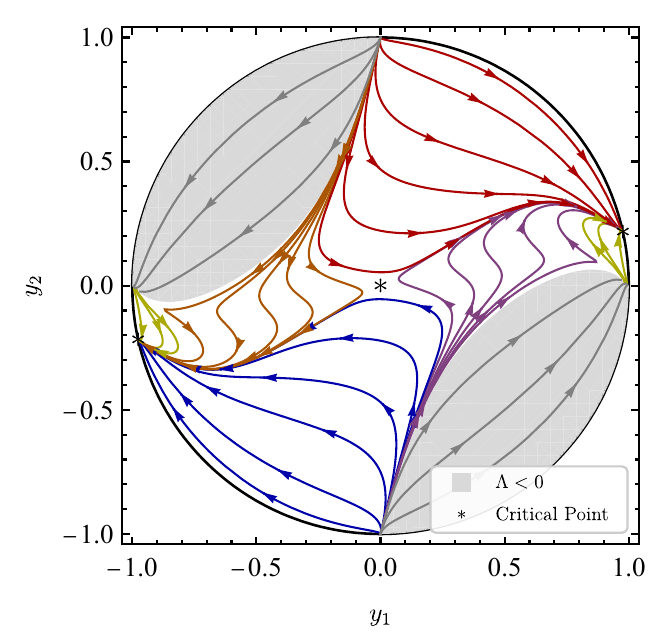} 
    \caption{\label{fig:phaseimg2} Global phase portrait of the dynamical system after Poincaré compactification. $y_1$ and $y_2$ are dimensionless quantities obtained by projecting $x_1$ and $x_2$ onto the Poincaré hemisphere. The asterisk at the origin denotes the unstable de Sitter solution (saddle point), while the asterisk on the boundary marks the universal attractor at infinity corresponding to the stiff-fluid epoch ($w_{\text{eff}}=+1$). Trajectories in the physical region ($\Lambda > 0$, unshaded) naturally evolve from the de Sitter phase towards the attractor, illustrating the elegant exit. The shaded grey regions indicate the unphysical regime where $\Lambda < 0$.}
\end{figure}

\begin{figure}[t] % [b] 表示尽量放在底部，也可以用 [t] 放在顶部
    % width=\columnwidth 是关键，它让图片宽度自动等于当前分栏的宽度
    \includegraphics[width=0.85\columnwidth]{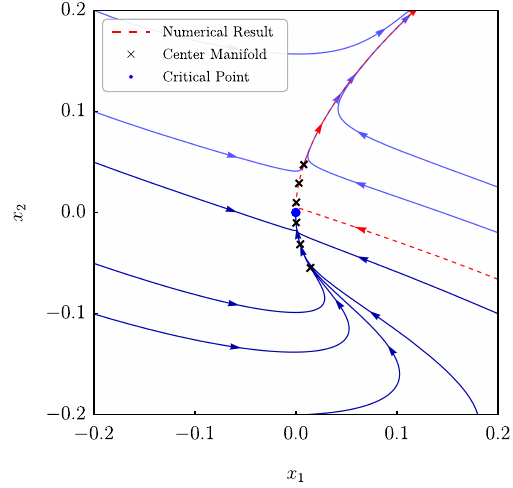} 
    \caption{\label{fig:phaseimg3} Dynamical behavior near the center manifold, the dynamical trajectories near the center manifold are rapidly attracted onto the center manifold marked by `$\times$' and then evolve along it. The red dashed line is a dynamical trajectory that achieves an e-folding number of 55, which corresponds to the red line in Fig. \ref{fig:Img1}}. 
\end{figure}

The origin $(0,0)$ represents an exact de Sitter solution with eigenvalues $\lambda_{1,2} = \left(-3 \pm \sqrt{9 + 32 \alpha_4/\alpha_1}\right)/2$. For $\alpha_4/\alpha_1 > 0$, this fixed point becomes a saddle, providing the necessary instability for the system to evolve away from the vacuum-dominated phase. Global analysis reveals a universal attractor at infinity, located at $({9}/{\sqrt{85}},{2}/{\sqrt{85}})$ with strictly negative eigenvalues, corresponding to a stiff-fluid epoch ($w_{\text{eff}} = +1$). The trajectory connecting this unstable saddle to the stable attractor naturally realizes an elegant exit, as visualized in the Poincaré phase portrait (FIG. \ref{fig:phaseimg2} with $\alpha_1 = \alpha_2 = \alpha_4 = 1$). Linearizing the dynamics near the saddle point allows us to derive the first slow-roll parameter in terms of $\Delta N \equiv N_{\text{end}} - N$. By introducing $\beta \equiv \sqrt{9 + 32 \alpha_4/\alpha_1}$, the expression simplifies to
\begin{equation}
    \epsilon_{H} = \frac{\alpha_1}{16} \beta (\beta - 3)^3 e^{\frac{3 - \beta}{2} \Delta N}.
\end{equation}

\subsection{Power-law slow-roll parameter}
While the exponential slow-roll model successfully realizes an elegant exit, it might demands fine-tuned initial conditions to sustain a sufficiently long inflationary period \cite{lindeProblemInitialConditions2018}. To mitigate this fine-tuning issue within the framework of $\Lambda$-driven inflation, we construct a second scenario that yields a power-law evolution of the slow-roll parameters. The central strategy involves modifying the phase space structure such that the de Sitter solution becomes a non-hyperbolic critical point. In this dynamical configuration, trajectories are rapidly attracted to a center manifold, along which they subsequently undergo a slow evolution. Relying on the dynamics of the center manifold not only alleviates the sensitivity to initial conditions but also naturally generates slow-roll parameters with power-law characteristics.

To realize the objective of constructing a non-hyperbolic critical point in phase space, we relax the restrictions on the functional forms of $V(\phi)$ and adopt a dimensionless scalar field. Within this framework, the potential functions are reformulated as products of dimensioned constants and dimensionless functions $f_i(\phi)$. A characteristic Hubble parameter $H_\mathrm{ds}$ with dimension $[T]^{-1}$ is introduced to set the physical scale, defined via the bare cosmological constant as $H_\mathrm{ds}^2 = \Lambda c^2/3$. In terms of this characteristic scale, the potential functions are parameterized as $V_1(\phi) = (c/H_\mathrm{ds})^2 f_1(\phi)$ and $V_4(\phi) = (c/H_\mathrm{ds})^2 f_4(\phi)$, with the remaining terms given by $V_2(\phi)=(c/H_\mathrm{ds})^4 f_2(\phi)$ and $V_3(\phi) = f_3(\phi)$. This parameterization ensures that $V_1$ and $V_4$ scale with the square of the characteristic length $c/H_\mathrm{ds}$, while $V_2$ scales with the fourth power, thereby preserving the dimensional consistency of the action.

It is worth noting that the cosmological dynamics of Fab-Four gravity in the FLRW background have been established by the original proponents of the theory. By applying the minisuperspace approximation to the action—varying it with respect to the homogeneous and isotropic metric ansatz—the generalized first Friedmann equation and the scalar field equation of motion are derived \cite{charmousisGeneralSecondOrderScalarTensor2012}. In this work, we adopt these established field equations as our starting point. Rather than solving for a specific potential, our strategy employs these equations to inversely constrain and select the functional forms of $f_i(\phi)$, guided by the dynamical objective of engineering a phase space structure with a non-hyperbolic critical point. Within our notation conventions, the modified Friedmann equation is expressed as a linear summation of the Fab-Four Hubble functions, $\sum_{i=1}^4 \mathcal{H}_i = -2\Lambda$. Correspondingly, the scalar field equation of motion is governed by the vanishing sum of the dynamical components, $\sum_{i=1}^4 \mathcal{E}_i = 0$. The explicit functional forms for the individual terms $\mathcal{H}_i$ and $\mathcal{E}_i$ are provided in the footnote \footnote{The Hubble functions are given by $\mathcal{H}_{1}=9c^{-4}V_{1}\dot{\phi}^{2}H^{2}$, $\mathcal{H}_{2}=-15c^{-6}V_{2}\dot{\phi}^{3}H^{3}$, $\mathcal{H}_{3}=-6c^{-2}(V_{3}H^{2}+V_{3}^{\prime}\dot{\phi}H)$, and $\mathcal{H}_{4}=-24c^{-4}V_{4}^{\prime}\dot{\phi}H^3$. The components of the scalar equation of motion are $\mathcal{E}_{1}=6(a^{3}c^{-4}V_{1}\dot{\phi}H^2)-3a^{3}c^{-4}V_{1}'\dot{\phi}^{2}H^2$, $\mathcal{E}_{2}=-9(a^{3}c^{-6}V_{2}\dot{\phi}^{2}H^3)+3a^{3}c^{-6}V_{2}'\dot{\phi}^{3}H^3$, $\mathcal{E}_{3}=-6c^{-2}(a^{3}V_{3}H)+6a^{3}c^{-2}H(V_{3}''\dot{\phi}+V_{3}'H)$, and $\mathcal{E}_{4}=-8V_{4}^{\prime}(a^{3}H^3)$, where the prime denotes the derivative with respect to scalar field $\phi$.}.

Since our scenario posits inflation fueled by the bare vacuum energy $\Lambda$, the dynamical system must admit an exact de Sitter solution as a critical point. This imposes the stationarity conditions $\dot{\phi} = 0$, $\ddot{\phi}=0$, and $\dot{H}=0$ when $H=H_\mathrm{ds}$. To engineer a critical point with the desired stability properties (specifically non-hyperbolicity), constraints must be placed on the potential functions $V(\phi)$. Building upon the equations of the $\mathcal{H}$ and the scalar equation of motion $\mathcal{E}$ derived in \cite{charmousisGeneralSecondOrderScalarTensor2012}, our strategy involves analyzing linear perturbations around the de Sitter solution to identify potentials that yield a vanishing eigenvalue in the Jacobian matrix.
For the perturbative analysis, we define the dimensionless variables $x_1 = \dot{\phi}/H_\mathrm{ds}$, $x_2 = H/H_\mathrm{ds}$, and $x_3 = \phi$. In this representation, the de Sitter solution corresponds to the fixed point $(x_1, x_2, x_3) = (0, 1, 0)$, where the scalar field value is set to zero . Substituting these variables into the generalized first Friedmann equation and the scalar field equation of motion while converting time derivatives to derivatives with respect to the $e$-folding number $N$ yields the evolution equations.

The resulting system consists of the constraint equation derived from the Friedmann equation,
\begin{equation}
    3 x_1^2 x_2^2 f_1 + 2 = 2 x_2 \left[ x_1 \left(4 x_2^2 f_4' + f_3'\right) + x_2 f_3 \right],
\end{equation}
and the dynamical equation of motion,
\begin{eqnarray}
    &&2 x_1 \left(3 x_2^3 f_2 \frac{\mathrm{d}x_1}{\mathrm{d}N} - 2 x_2 f_1 \frac{\mathrm{d}x_2}{\mathrm{d}N} - 3 x_2^2 f_1 + f_3''\right) \nonumber \\
    &&+ x_2^2 \left(8 f_4' \frac{\mathrm{d}x_2}{\mathrm{d}N} - 2 f_1 \frac{\mathrm{d}x_1}{\mathrm{d}N}\right) \nonumber \\
    &&+ x_1^2 x_2 \left(9 x_2 f_2 \frac{\mathrm{d}x_2}{\mathrm{d}N} + 9 x_2^2 f_2 - f_1'\right) \nonumber \\
    &&+ 2 x_1^3 x_2^2 f_2' + 2 f_3' \frac{\mathrm{d}x_2}{\mathrm{d}N} + 4 x_2 f_3' + 8 x_2^3 f_4' = 0,
\end{eqnarray}
where we have suppressed the explicit dependence of $f_i$ on $\phi$ (or $x_3$) for brevity, and primes denote derivatives with respect to the scalar field. Substituting the dynamical variables corresponding to the de Sitter critical point, $(x_1, x_2, x_3) = (0, 1, 0)$, into the system equations, consistency requires the potential functions to satisfy the following constraints 
\begin{equation}
    \begin{cases}
        f_3'(0)+2f_4'(0)=0, \\
        f_3(0) = 1.
    \end{cases}
\end{equation}
Performing a perturbative analysis around this critical point, it can be rigorously demonstrated that the de Sitter solution manifests as a non-hyperbolic critical point, characterized by a vanishing eigenvalue in the Jacobian matrix, provided that the second derivatives satisfy the condition of
\begin{equation}
    f_3''(0) + 2 f_4''(0) = 0.
\end{equation}
Assuming the standard Einstein-Hilbert term is retained in $\mathcal{L}_3$ and considering a power-law ansatz $f_4(\phi) = \alpha_4 \phi^n$, the exponent is constrained to satisfy $n>2$. Adopting the lowest integer power leads to $V_4(\phi) = \alpha_4 (c/H_\mathrm{ds})^2 \phi^3$, which, combined with the potentials $V_1(\phi) = \alpha_1(c/H_\mathrm{ds})^2$ and $V_2(\phi) =  \alpha_2 (c/H_\mathrm{ds})^4 \phi^2$, constitutes Model II. An analysis of the critical points at infinity, analogous to that performed for Model I, confirms that this model shares the same asymptotic behavior, yielding $w_\text{eff} \to +1$.

To investigate the non-hyperbolic critical point central to this model, a distinct set of dimensionless variables is employed: $x_1 = \dot{\phi}/H_\mathrm{ds}$, $x_2 = \phi\sqrt{H_\mathrm{ds}/H}$, and $x_3 = H_\mathrm{ds}/H$. Here, $x_3$ does not constitute an independent dynamical variable, as it is algebraically constrained by the first Friedmann equation. The phase portrait derived from the flow equations $\mathrm{d}x_1/\mathrm{d}N$ and $\mathrm{d}x_2/\mathrm{d}N$ reveals this critical point to be a saddle-node. Consequently, standard linearization techniques based on the Hartman-Grobman theorem break down, necessitating a nonlinear analysis via Center Manifold Theory\cite{meissDifferentialDynamicalSystems2007,2018PhR...775....1B}. As illustrated in the FIG. \ref{fig:phaseimg3}, trajectories are rapidly attracted to the center manifold emerging from the critical point, followed by a slow evolution along it. Given that the local behavior is dominated by the dynamics on the center manifold, an effective reduction in dimensionality is achieved by restricting the system to this manifold, significantly simplifying the analysis of the nonlinear flow.

Model II constitutes a two-dimensional system subject to the algebraic constraint of the first Friedmann equation. While the constraint is too complex to yield a global closed-form expression for $x_3(x_1,x_2)$, it can be solved perturbatively in the vicinity of the critical point. This local approximation of the constraint allows for the calculation of the center manifold as a power series expansion around the fixed point. Its leading term gives $x_1 = 4 x_2^2$,
whose trajectory is depicted in the FIG. \ref{fig:phaseimg3} marked by `$\times$'. 

By restricting the analysis to the dynamics on this computed manifold, the evolution of physical quantities near the critical point can be determined analytically. Consequently, the first slow-roll parameter for Model II is found to have the approximate form
\begin{equation}
    \epsilon_{H} \approx \frac{3}{16}\left(\frac{\alpha_1}{\alpha_4}\frac{1}{\Delta N}\right)^5.
\end{equation}
This result establishes a power-law behavior for the slow-roll parameter, which significantly alleviates the fine-tuning of initial conditions required by its exponential counterpart. In this scenario, phase space trajectories are rapidly channeled into the basin of attraction of the center manifold \cite{2018PhR...775....1B}, akin to falling into a valley, and subsequently evolve slowly along its floor, thereby sustaining a prolonged period of inflation.

Finally, Fig. 3 displays the numerical evolution of $\epsilon_H$ for both models. With appropriate initial conditions, an inflationary duration of $N \approx 55$ is readily achieved using parameters $\alpha_1=20, \alpha_2=\alpha_4=1$ for Model I and $\alpha_1=\alpha_2=\alpha_4=1$ for Model II. The dashed lines indicate the corresponding analytical approximations near the de Sitter solution. Notably, Model I exhibits a distinct intermediate plateau deviating from the exponential approximation. This feature arises from non-linear higher-order derivative couplings that create a dynamical "bottleneck," effectively suppressing the growth of $\epsilon_H$, prolonging the quasi-de Sitter phase and and it may lead to unique observational effects.

\begin{figure}[t]
\includegraphics[width=\columnwidth]{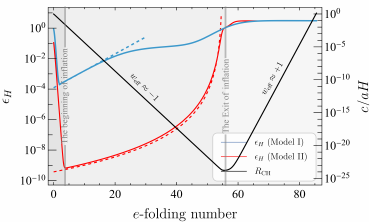}% Here is how to import EPS art
\caption{\label{fig:Img1} Evolution of the first slow-roll parameter  $\epsilon_H$ (left axis, red line and blue line) and the comoving Hubble Radius $R_\mathrm{CH}=c/(aH)$ (right axis, solid black line) as a function of the e-folding number $N$ for these models. Inflation proceeds while $\epsilon_H \ll 1$ and terminates when $\epsilon_H = 1$, marking a smooth transition from a quasi-de Sitter phase to a stiff-fluid dominated epoch. The dashed line shows the excellent agreement between our analytic approximation for $\epsilon_H$ and the full numerical solution. }
\end{figure}

\section{Conclusion} In this Letter, we have presented two explicit Fab-Four constructions in which a de Sitter phase sourced by a bare cosmological constant is not the endpoint of the dynamics but an unstable critical point. The subsequent evolution toward a stiff-fluid attractor provides a graceful exit from vacuum-energy-driven inflation. In Model I the exit is controlled by an ordinary saddle and the first slow-roll parameter grows exponentially. In Model II the de Sitter point is non-hyperbolic; trajectories are rapidly attracted to a center manifold and then drift slowly along it, leading to a power-law growth of $\epsilon_H$ and reducing the sensitivity to initial conditions.

Because the post-inflationary attractor is kination-like, reheating cannot proceed through the usual oscillatory mechanism. Gravitational particle production, as in quintessential inflation \cite{parkerProductionElementaryParticles1977,peeblesQuintessentialInflation1999}, is a plausible alternative, but its quantitative viability depends on the efficiency of the de Sitter-to-kination transition. We do not address these issues here. A more basic limitation is that the present Fab-Four realization is not selective: the same self-tuning structure that screens vacuum energy also tends to screen matter and radiation, obstructing the standard Friedmann era. Nevertheless, this indiscriminate screening is merely a shortcoming of the present implementation, not necessarily of self-tuning more generally \cite{applebyWellTemperedCosmologicalConstant2018,khanMinimalSelfTuning2022}. The inflationary scenario proposed here is readily compatible with other self-tuning constructions.
%In addition, derivative couplings of this kind are tightly constrained by the luminal propagation of gravitational waves and by perturbative stability requirements.

Summarizing, the constructions presented in this Letter should be viewed as proofs of principle rather than complete cosmological models. Their main lesson is that inflation sourced by a bare cosmological constant and a graceful exit from it are dynamically compatible. The remaining challenge is to implement this mechanism in a selective self-tuning framework that preserves standard cosmology and satisfies gravitational-wave and stability constraints. A full perturbative analysis, including the primordial spectra, is deferred to future works.

\section{Acknowledgements} This work is supported by National Natural Science Foundation of China under Grants Nos. 12021003, 12433001, 12405050; and Beijing Natural Science Foundation No. 1242021.

\bibliography{Lambda,excite}% Produces the bibliography via BibTeX.

\end{document}